\documentclass[aps,amsmath,amssymb,prl,10pt,twocolumn,superscriptaddress,]{revtex4-1}
\usepackage{graphicx,txfonts}
\usepackage{bm}
\usepackage{mathrsfs}
\usepackage{comment}
\usepackage{mathtools}
\usepackage{appendix}
\usepackage{color}
\usepackage{hyperref}
\usepackage{bbm}
\usepackage{float}
\usepackage[caption = false]{subfig}

\newcommand{\al}{\alpha}
\newcommand{\be}{\beta}

\newcommand{\de}{\delta}

\newcommand{\la}{\lambda}

\newcommand{\p}{\pi}

\newcommand{\s}{\sigma}

\newcommand{\x}{\chi}
\newcommand{\w}{\omega}
\newcommand{\W}{\Omega}

\newcommand{\ta}{\tau}

\newcommand{\round}[1]{\left( #1 \right)}
\renewcommand{\square}[1]{\left[ #1 \right]}

\newcommand{\ang}[1]{\left\langle #1 \right\rangle}

\newcommand{\beq}{\begin{equation}}
\newcommand{\eeq}{\end{equation}}
\newcommand{\Beq}{\begin{eqnarray}}
\newcommand{\Eeq}{\end{eqnarray}}
\newcommand{\bml}{\begin{multline}}

\newcommand{\bea}{\begin{align}}
\newcommand{\ena}{\end{align}}
\newcommand{\bsp}{\begin{split}}
\newcommand{\esp}{\end{split}}
\newcommand{\down}{\downarrow}
\newcommand{\up}{\uparrow}

\newcommand{\ex}{\hat{\boldsymbol x}}

\newcommand{\bi}{{\boldsymbol i}}
\newcommand{\bj}{{\boldsymbol j}}

\newcommand{\bk}{{\boldsymbol k}}

\newcommand{\sA}{\mathscr{A}}
\newcommand{\sN}{\mathscr{N}}

\newcommand{\hH}{\hat{H}}

\newcommand{\hS}{\hat{S}}

\newcommand{\hbS}{\hat{\boldsymbol{S}}}
\newcommand{\hbs}{\hat{\boldsymbol{s}}}

\newcommand{\bx}{\boldsymbol{x}}
\newcommand{\bI}{\boldsymbol{I}}

\newcommand{\hc}{\hat{c}}
\newcommand{\ve}{\varepsilon}

\newcommand{\bta}{{\boldsymbol{\tau}}}

\newcommand{\hs}{\hat{s}}
\newcommand{\htt}{\hat{t}}

\newcommand{\hI}{{\hat I}}

\newcommand{\bR}{{\boldsymbol R}}

\newcommand{\hT}{\hat{T}}

\newcommand{\hy}{\hat{\psi}}

\newcommand{\bsig}{{\boldsymbol \sigma}}

\newcommand{\jp}{J} % Inter-chain coupling Jp
\newcommand{\jn}{K} % Intra-chain NN Heisenberg coupling 
\newcommand{\dm}{D} % Intra-chain DM interaction
\newcommand{\gm}{\Gamma} % Bond anisotropy
\newcommand{\hhy}{h_y} % External field

\newcommand{\tx}{\htt_{x}}
\newcommand{\txd}{\htt^{\dagger}_{x}}
\newcommand{\ty}{\htt_{y}}
\newcommand{\tyd}{\htt^{\dagger}_{y}}
\newcommand{\tz}{\htt_{z}}
\newcommand{\tzd}{\htt^{\dagger}_{z}}

\newcommand{\uy}{\bar{u}}
\newcommand{\vy}{\bar{v}}

\newcommand{\ii}{\iota} % Math symbol for \sqrt(-1)
\begin{document}
\title{Detecting End-States of   Topological Quantum Paramagnets via Spin Hall Noise Spectroscopy}

\date{\today}
\author{Darshan G. Joshi}
\author{Andreas P. Schnyder}
\affiliation{Max-Planck-Institute for Solid State Research, Heisenbergstrasse 1, 70569 Stuttgart, Germany}
\author{So Takei}
\affiliation{Department of Physics, Queens College of the City University of New York, Queens, NY 11367, USA}
\affiliation{Physics Doctoral Program, The Graduate Center of the City University of New York, New York, NY 10016, USA}

\begin{abstract}
We theoretically study the equilibrium spin current fluctuations and the corresponding charge noise generated by inverse spin Hall effect (ISHE) in a metal with strong spin-orbit coupling deposited on top of a quantum paramagnet.
 It is shown that the charge noise power spectra measured along different spatial axes can  directly probe the different spin components of the boundary dynamic spin correlations of the quantum paramagnet. We report the utility of this ISHE-facilitated spin noise probe as a tool to unambiguously detect topological phase transitions in an $S=1/2$ quantum spin ladder that hosts a trivial ground state of singlet product states, but topologically-protected fractional spin excitations localized at its ends. Our work demonstrates the general usefulness of the ISHE-mediated spin noise spectroscopy for the detection of topological phases in quantum paramagnets.
\end{abstract}
\maketitle

%%%%%%%%%%%%%%%%%%%%%%%%%%%%%%%%%%%%%%%%%%%%%%%%%%%%%%%%%%%%%%%%%%%%
For over the last decade, spin noise spectroscopy has provided a powerful tool to study the dynamics of interacting spin systems through their spin fluctuations~\cite{aleksandrovJPCO11,*zapasskiiAOP13,*huebnerPSS14,*sinitsynRPP16}. A well-studied mode of operation is the optical approach~\cite{aleksandrovJETP81}, in which these fluctuations are quantified using fluctuations in the Faraday rotation angle for a linearly polarized beam passing through the sample. Inverse spin Hall effect (ISHE) refers to a relativistic spin-orbit coupling phenomenon, in which a pure spin current flowing in a paramagnetic conductor converts into a transverse charge current~\cite{maekawaBOOK12,*hoffmannIEEEM13,*sinovaRMP15,*niimiRPR15}. The effect has been extensively used in spintronics as an essential tool for the detection of pure spin currents via electrical signals~\cite{saitohAPL06,*valenzuelaNAT06,*zhaoPRL06,*andoPRL08,*mosendzPRB10,*andoJAP11,*chumakAPL12,*hahnPRB13,*jungfleischPRB15}. Exploiting the utility of ISHE as a spin-to-charge transducer, it is interesting to explore how spin noise spectroscopy can be performed on interacting spin systems by converting the spin noise information into charge noise signals via the ISHE.

Such ISHE-facilitated spin current noise spectroscopy, referred to here as spin Hall noise spectroscopy (SHNS), has recently attracted attention in bilayer systems consisting of a normal metal in contact with a quantum magnet with long-range magnetic order~\cite{kamraPRB14,kamraPRL16,matsuoPRL18}. Spin fluctuations in the quantum magnet lead to spin current fluctuations near the interface that diffuse into the metal. These fluctuations can eventually be converted into charge fluctuations via the ISHE and ultimately detected electrically. Theoretical works have shown that SHNS can be used to reveal the quantum uncertainty associated with magnon eigenstates~\cite{kamraPRL16} as well as non-trivial spin scattering and heating processes taking place at such an interface~\cite{matsuoPRL18}. A recent experimental work investigated equilibrium charge noise in Pt thin films deposited on top of an insulator with long-ranged ferromagnetic order~\cite{kamraPRB14}. It was observed that the equilibrium voltage noise power spectrum measured across the Pt film depends on the magnetization orientation~\cite{kamraPRB14} in a nontrivial way. This result was interpreted as a modulation in the thermal spin current noise in the metal, due to variations in the magnetization orientation, which led to modulations in the charge noise via the ISHE. 

\begin{figure}[t]
\centering
\includegraphics[width=0.38\textwidth]{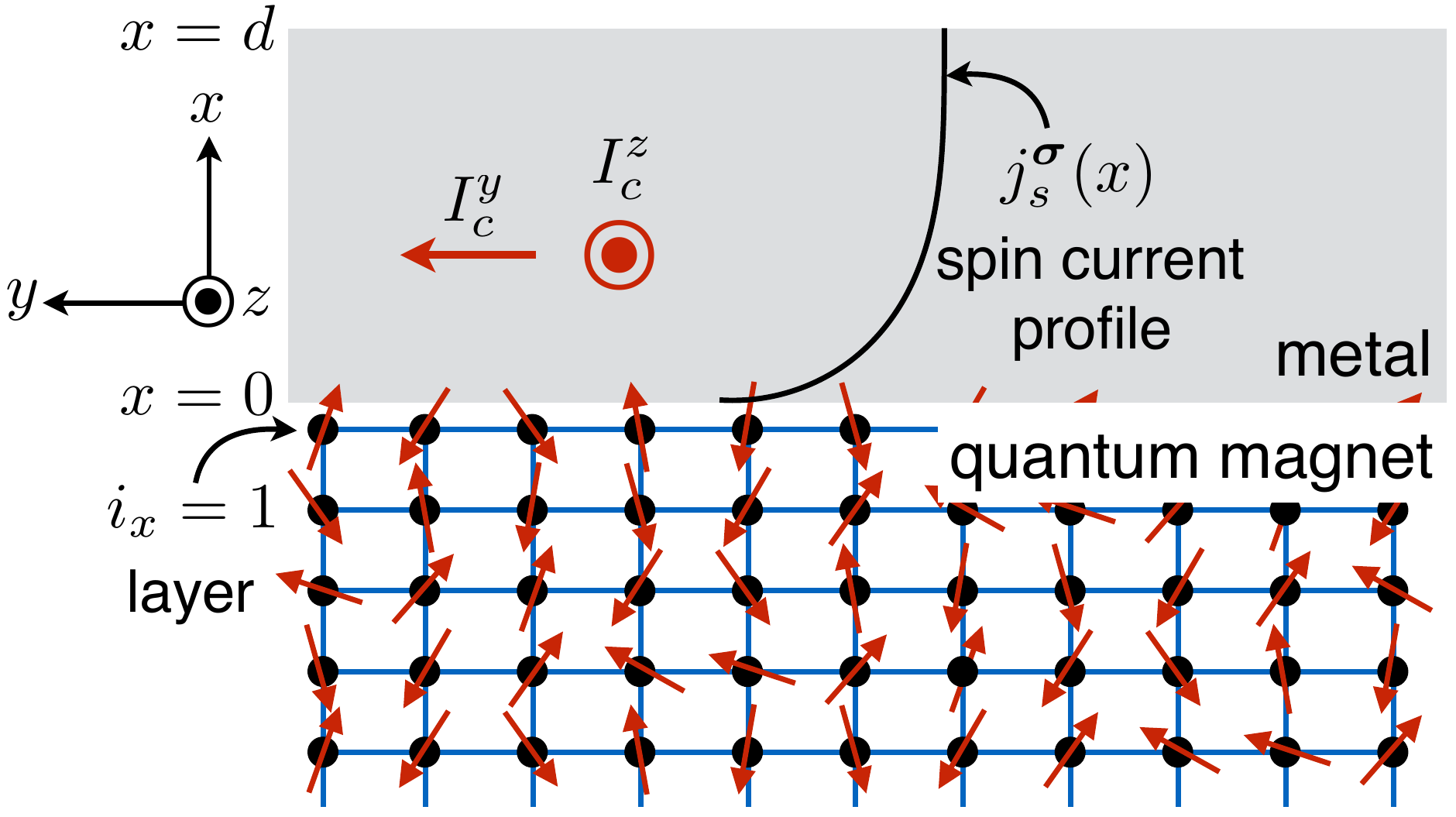}
\caption{Schematic diagram of the setup for spin Hall noise spectroscopy. A quantum magnet is coupled at its top surface to a normal metal film with strong spin-orbit coupling. Spin fluctuations in the quantum magnet lead to a fluctuating pure spin current in the metal and to a fluctuation charge current via the inverse spin Hall effect.}
\label{fig1}
\end{figure}

In this work, we consider SHNS in a bilayer consisting of a normal metal in contact with a quantum paramagnet, which is a quantum magnet with quantum fluctuations strong enough to destroy conventional magnetic ordering~\cite{lhuillierBOOK01,*richterBOOK04,*sachdevBOOK04,*balentsNAT10}~(see Fig.~\ref{fig1}). In particular, we report the utility of the SHNS setup as a tool to unambiguously detect topological phase transitions in such a quantum paramagnet. Following the advent of fermionic systems with nontrivial symmetry-protected topological order~\cite{hasanRMP10,*qiRMP11,*chiuRMP16}, the emergence of elementary bosonic excitations with topologically nontrivial band structures has been investigated in the context of various quantum spin models~\cite{romhanyiNATC15,liNATC16af,mcclartyNATP17,joshiCM18,mcclartyCM18,joshiPRB17}. Akin to the fermionic systems, the nontrivial  topology of these bosonic bands gives rise to protected magnon or triplon edge states, which distinguish a topological quantum paramagnet from its trivial counterpart. We demonstrate the suitability of SHNS as a means to detect topological phase transitions in quantum paramagnets, by considering as a concrete example  an $S=1/2$ quantum spin ladder hosting a topological quantum paramagnet (TQP)~\cite{joshiPRB17}. The TQP considered here is an exotic state of matter with a trivial ground state of singlet product states, but hosts fractional excitations localized at the ends. Using SHNS we show that we can access the dynamical spin correlations at the ends, which bear the signatures of these non-trivial end states. 

We begin with a general discussion of SHNS. An attractive aspect of SHNS is that it allows to use a table-top experiment to probe dynamical spin correlations for virtually any type of quantum spin system, by depositing above it a normal metal film with strong spin-orbit coupling (e.g., Pt, Ta, W, etc., as shown in Fig.~\ref{fig1}). Moreover, the experiment can be performed in thermal equilibrium, thus eliminating any unwanted effects (such as Joule heating and shot noise) that might arise in the presence of nonequilibrium drives. We assume that the entire structure is in thermal equilibrium at temperature $T$ and, for simplicity, consider a quantum magnet with a cubic lattice structure with quantum spins $\hbS_\bi$ localized on the lattice sites $\bi$. The exact lattice structure is not germane to the rest of the discussion. 

The quantum magnet and the normal metal are assumed to be coupled via an isotropic exchange interaction,
\beq
\label{hc}
\hH_c=-\eta a^3\sum_{\bi}\de_{i_x,1}\hbs(x=0,\bR_\bi)\cdot\hbS_\bi\ ,
\eeq
where $\eta$ is the exchange constant and $a$ is the lattice scale of the metal. Here, $\hbs(\bx)=\hy^\dag_{s}(\bx)\bta_{ss'}\hy_{s'}(\bx)/2$ is the local spin density in the metal, 
 $x=0$ is the interface plane of the metal, and  $\bR_\bi$ collects the 2d interfacial coordinates corresponding to lattice sites $\bi$ on the $i_x=1$ (interfacial) layer (see Fig.~\ref{fig1}),

Spin fluctuations in the quantum magnet generate a fluctuating spin current $\bI^\bsig_s(t)=I^\bsig_s(t)\ex$ with spin polarization $\bsig$ in the normal metal at $x=0$. For an isotropic diffusive metal with spin diffusion length $\la$ (and for frequencies much smaller than the inverse spin relaxation time in the metal), the spin current density profile inside the metal may be written as~\cite{mosendzPRB10,kamraPRB14}
\beq
\label{js}
\bj^\bsig_s(x,t)=\bI^\bsig_s(t)\frac{\sinh[(d-x)/\la]}{\sA_i\sinh(d/\la)}\ ,
\eeq
where $d$ is the thickness of the normal metal film, $\sA_i$ is the interfacial area, and a boundary condition of vanishing spin current at the outer edge is assumed. Via the ISHE, the fluctuating spin current Eq.~\eqref{js} leads to charge current fluctuations (integrated over the cross-sectional area $\sA_m$ normal to charge flow direction), given by
\beq
\label{ic}
\bI_c(t)=\Theta\frac{2e}{\hbar}\frac{\la\sA_m}{d\sA_i}\tanh\round{\frac{d}{2\la}}(\ex\times\bsig)I^\bsig_s(t)\ ,
\eeq
where $\Theta$ is the so-called spin Hall angle. As a consequence, the charge noise spectrum measured along the $y$ ($z$) axis [denoted by $S^{y,z}_c(\W)$] is sensitive to the spectrum of spin current fluctuations polarized along the $z$ ($y$) axis [denoted by $S^{z,y}_s(\W)=\int dt\ \langle I^{z,y}_s(0)I^{z,y}_s(t)\rangle e^{-i\W t}$], i.e., 
\beq
\label{sc}
S^{y,z}_c(\W)=\square{\Theta\frac{2e}{\hbar}\frac{\la\sA_m}{d\sA_i}\tanh\round{\frac{d}{2\la}}}^2S^{z,y}_s(\W)\ .
\eeq

We now compute the $z$ component of the charge noise spectrum using the above equation. (The derivation for the $y$ component is similar and will be simply stated at a later point.) Orienting the spin quantization axis in the metal along the  axis of the spin polarization $\bsig$ in the quantum magnet, the  $y$ component of the operator for the spin current entering the metal reads $\hI_s^y=\ii\eta a^3\hT_y/2+h.c.$, where $\hT_y=\sum_{\bi}\de_{i_x,1}\hy^\dag_\up(\bR_\bi)\hy_\down(\bR_\bi)(\hS^z_\bi-\ii\hS^x_\bi)$,
and $\ii = \sqrt{-1}$.
%$ \hT_\bz&=\sum_{\bi}\de_{i_x,1}\hy^\dag_\up(\bR_\bi)\hy_\down(\bR_\bi)\round{\hS^x_\bi-i\hS^y_\bi}\ . $
For a metal with quadratic dispersion $\ve_\bk=\hbar^2k^2/2m$ and chemical potential $\mu$, the noise spectral density $S^y_s(\W)$ can then be computed to lowest non-trivial order in $\eta$, 
\begin{align} \label{noise_spectral_density}
S^y_s(\W)&=2\ii\round{\frac{\eta a^3mk_F}{2\p^2\hbar}}^2\sum_{\bi,\bj}\de_{i_x,1}\de_{j_x,1}\int d\nu\square{ \x^{xx}_{\bi\bj}(\nu) + \x^{zz}_{\bi\bj}(\nu)  } \nonumber\\
&\qquad\qquad\qquad\times{\rm sinc}^2(k_F|\bR_\bi-\bR_\bj|)\frac{\W-\nu}{e^{\be\hbar(\W-\nu)}-1}\ ,
\end{align}
where $\beta$ is the inverse temperature, $k_F$ is the Fermi wavevector in the metal, and the spin correlation functions in the quantum magnet are defined via
\beq \label{correlator_FT}
-\ii\ang{\hS^{\al}_{\bi}(0)\hS^{\al}_{\bj}(t)}_0=\int\frac{d\nu}{2\p}\ \x^{\al\al}_{\bi\bj}(\nu)e^{\ii\nu t} ,
\eeq
see supplemental material  (SM) for technical details of the derivation~\cite{supp_info}. For large Fermi wavevectors, i.e., $k_F|\bR_\bi-\bR_\bj|\gg1$ for all $\bi\ne\bj$, and in the low temperature limit, one may finally show that the second derivative of Eq.~\eqref{sc} reduces to
\begin{multline}
\label{d2sc}
\frac{d^2S^z_c(\W)}{d\W^2}=\square{\Theta\frac{2e}{\hbar}\frac{\la\sA_m}{d\sA_i}\tanh\round{\frac{d}{2\la}}\round{\frac{\eta a^3mk_F}{2\p^2\hbar}}}^2\\
\times 2\ii\sum_{\bi}\de_{i_x,1}\square{ \x^{xx}_{\bi\bi}(\W) + \x^{zz}_{\bi\bi}(\W)}\ .
\end{multline}
We see that the second derivative of the charge noise spectrum measured along the $z$ axis is directly proportional to the $x$ and $z$ components of the interfacial dynamical spin correlations of the quantum paramagnet. A similar calculation for the charge current fluctuations along the $y$ axis gives
\begin{multline}
\frac{d^2S^y_c(\W)}{d\W^2}=\square{\Theta\frac{2e}{\hbar}\frac{\la\sA_m}{d\sA_i}\tanh\round{\frac{d}{2\la}}\round{\frac{\eta a^3mk_F}{2\p^2\hbar}}}^2\\
\times 2\ii\sum_{\bi}\de_{i_x,1}\square{\x^{xx}_{\bi\bi}(\W)+\x^{yy}_{\bi\bi}(\W)}\ .
\end{multline}
Since the spin correlation functions are extracted via the frequency derivatives of the noise spectra, this detection method has the advantage of being able to eliminate any unwanted (frequency-independent) background white noise, e.g., Johnson-Nyquist noise.

We now discuss how the SHNS can be used to probe a TQP. The TQP considered here is a topologically non-trivial state of a quantum spin ladder system hosting trivial spin-1 excitations in the bulk and fractional spin excitations localized at the ends of the ladder~\cite{joshiPRB17}. The topological aspect in this case is manifested in the excitations, in contrast to the fermionic topological phases wherein the ground state carries the topological features, thus rendering the detection of a TQP challenging. We now discuss how the SHNS setup introduced above can provide a definitive experimental signature of the TQP. 

\begin{figure}[t]
\centering
\includegraphics[width=0.3\textwidth]{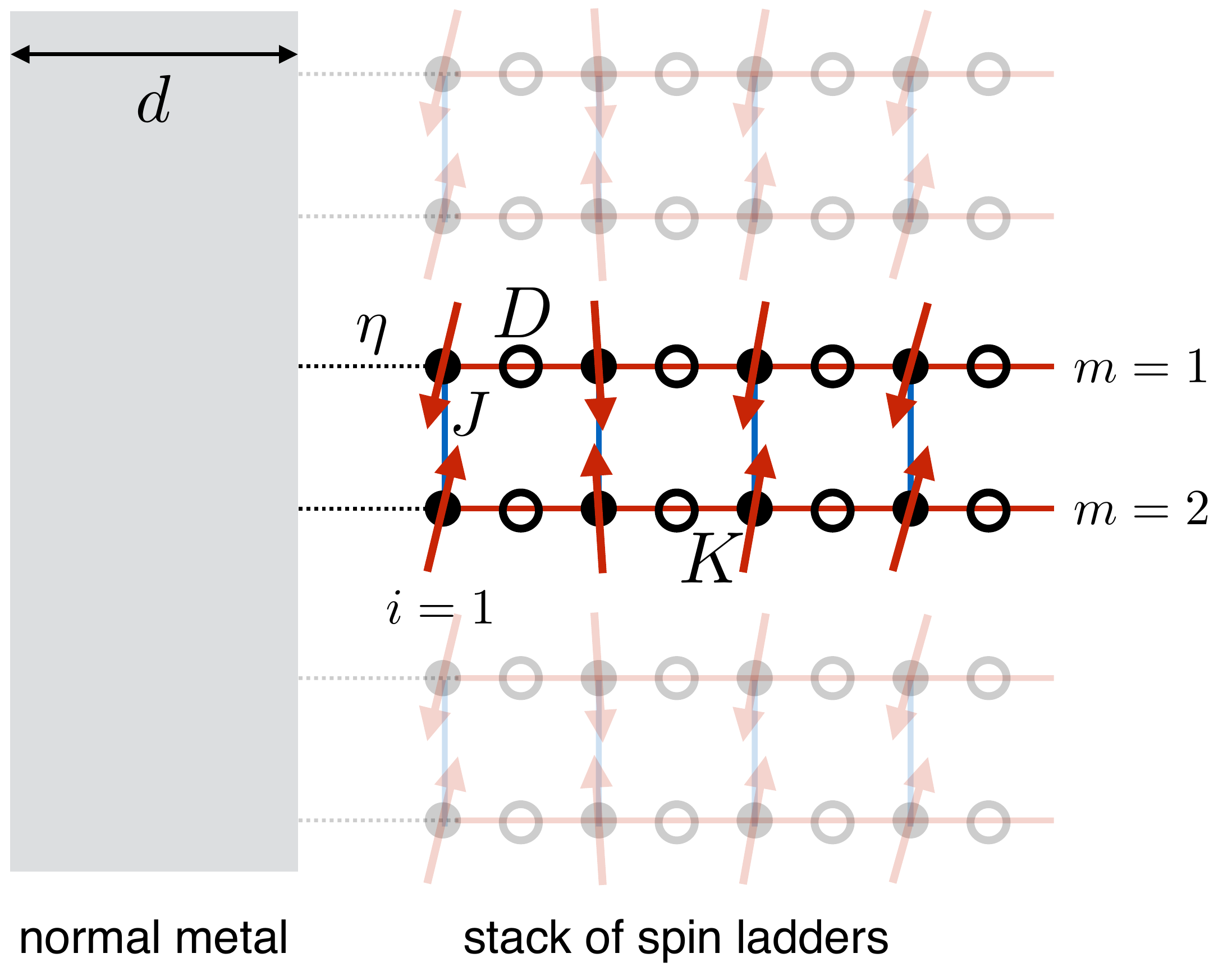}
\caption{A schematic diagram of a TQP (i.e., a quantum spin ladder) exchange-coupled to a normal metal at one of its ends.}
\label{fig2}
\end{figure}

The setup of interest is shown in Fig.~\ref{fig2}, in which $\sN$ quantum spin ladders (per unit interfacial area) are laterally exchange-coupled to a spin-orbit coupled metal at one of their ends. Since SHNS probes the local dynamical spin correlations at the ends of the ladders, the charge noise spectrum inherits the signatures of the topologically non-trivial end states and facilitates the electrical detection of the topological phase transition. 

The quantum spin ladder, hosting a TQP, is described by the following Hamiltonian~\cite{joshiPRB17}
\begin{align} \label{ham_TQP}
\hH&=\jp \sum_{i} \hbS_{1i} \cdot \hbS_{2i}+\jn \sum_{i} \big[ \hbS_{1i} \cdot \hbS_{1i+1} + \hbS_{2i} \cdot \hbS_{2i+1} \big] \nonumber \\
&+ \dm \sum_{i} \big[ \hS^{z}_{1i} \hS^{x}_{1i+1} - \hS^{x}_{1i} \hS^{z}_{1i+1}+ \hS^{z}_{2i} \hS^{x}_{2i+1} - \hS^{x}_{2i} \hS^{z}_{2i+1} \big] \nonumber \\
&+ \gm \sum_{i} \big[ \hS^{z}_{1i} \hS^{x}_{1i+1} + \hS^{x}_{1i} \hS^{z}_{1i+1}+ \hS^{z}_{2i} \hS^{x}_{2i+1} + \hS^{x}_{2i} \hS^{z}_{2i+1} \big]   \nonumber \\
&+ \hhy \sum_{i} \big[ \hS^{y}_{1i} + \hS^{y}_{2i} \big]\  ,
\end{align}
where  $i$ denotes the dimer site, $m=1,2$ labels the two legs of the ladder, $\jp$ is the antiferromagnetic intra-dimer coupling, and $\jn$ is the inter-dimer Heisenberg interaction. The odd-parity Dzyaloshinkii-Moriya (DM) interaction $\dm$ and the even-parity spin-anisotropic inter-dimer coupling $\gm$ arise from spin-orbit coupling. The TQP obtains for $|\hhy|<\dm$.

In the quantum paramagnetic phase, we represent the spins via bosonic quasiparticles, i.e., {\em triplons}, described within the bond-operator theory as follows~\cite{sachdevPRB90,romhanyiPRB11}
\begin{equation}
\label{eq:s_tri}
\hS^{\alpha}_{1,2i} = \frac{\ii}{2} \left( \pm  \htt^{\dagger}_{i\alpha} \hs_{i} \mp \hs^{\dagger}_{i} \htt_{i\alpha} - \epsilon_{\alpha\beta\gamma} 
\htt^{\dagger}_{i \beta} \htt_{i \gamma} \right) \,, %~~~~
%P_{i} = 1 - \sum_{\gamma} t^{\dagger}_{i\gamma} t_{i\gamma} \,,
\end{equation}
where $\htt$ ($\htt^{\dagger}$) are the triplon annihilation (creation) operators, coming in three flavors corresponding to the three triplet states. In the quantum paramagnetic phase, we condense the singlet operator $\hs$, such that we can replace $\hs=\hs^{\dagger}=1$, and within the harmonic approximation we retain only the bilinear terms in the triplon operators.

%%%%%%%%%%%%%%%%%%%%%%%%%%%%%%%%% Img-part plots (combo)
\begin{figure}
\centering
\subfloat{\includegraphics[width=0.24\textwidth]{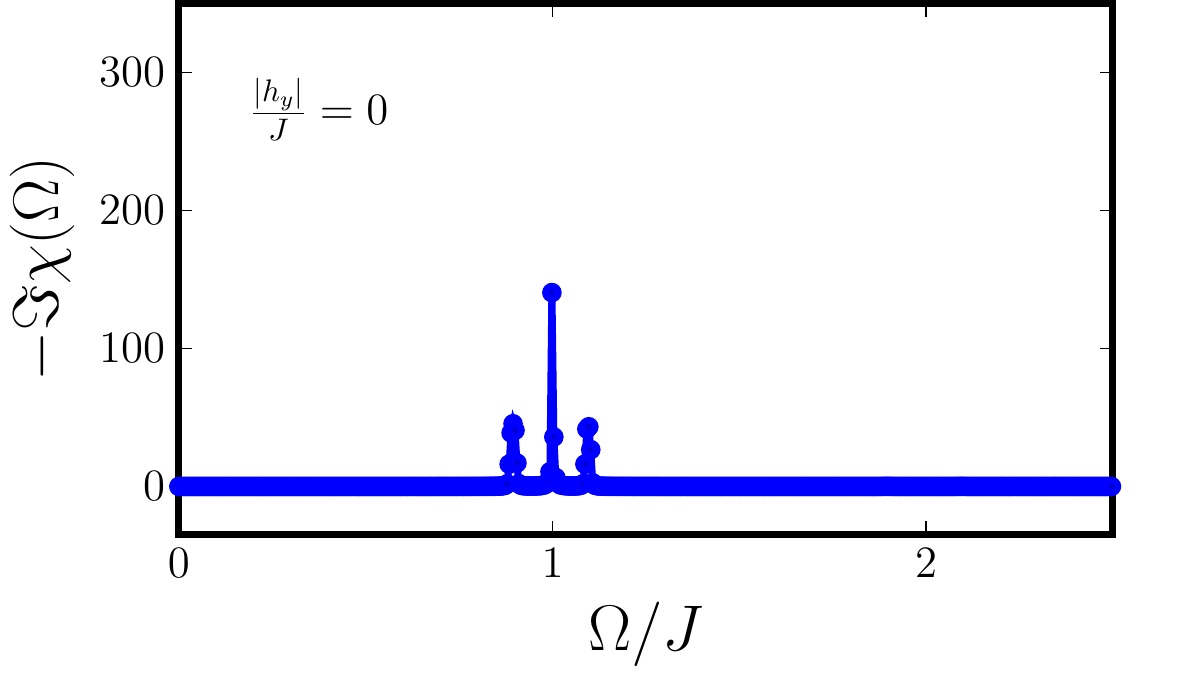}} ~~
\subfloat{\includegraphics[width=0.24\textwidth]{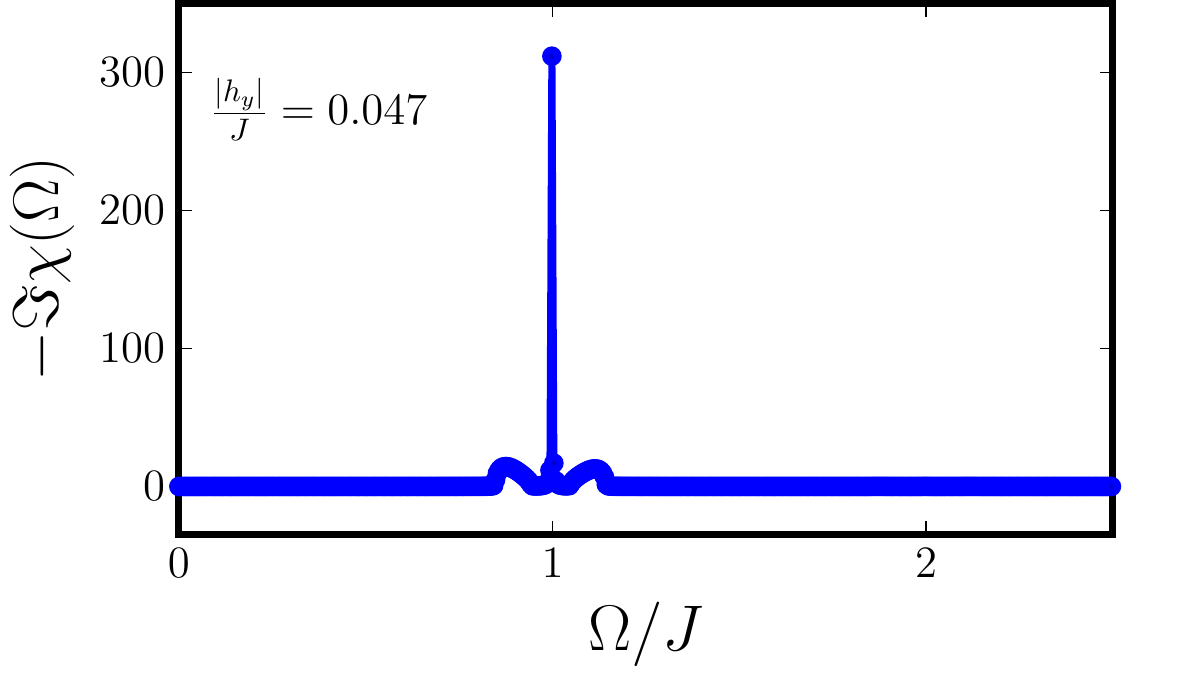}} \\
\subfloat{\includegraphics[width=0.24\textwidth]{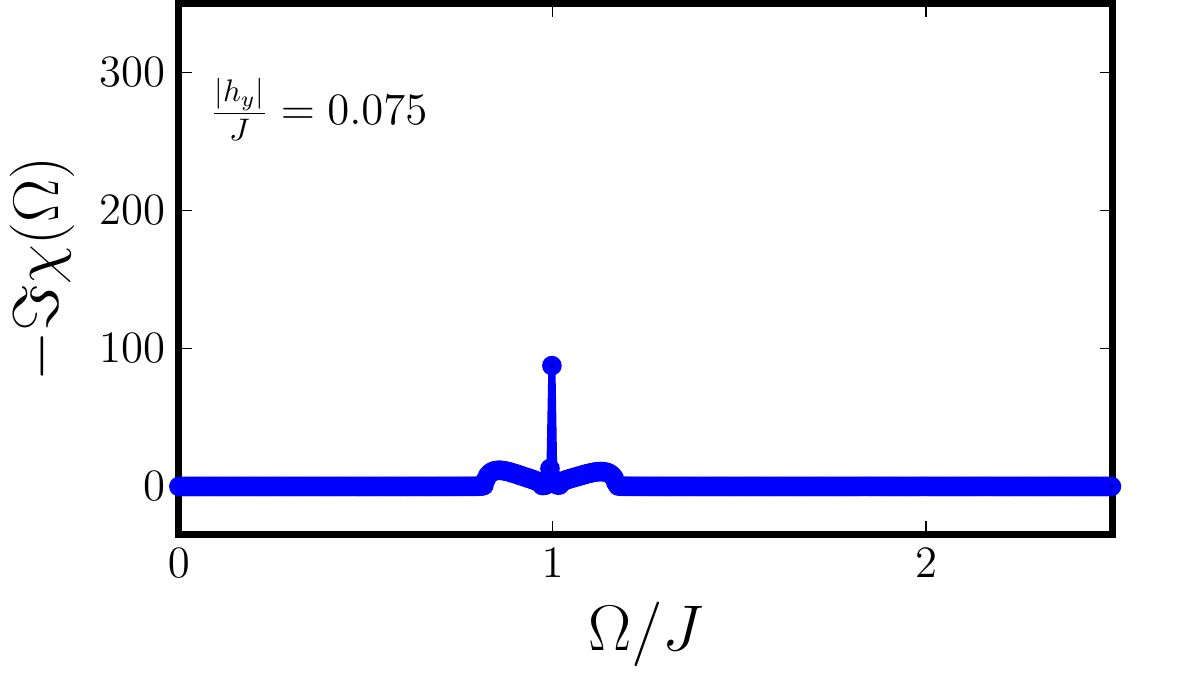}} ~~
\subfloat{\includegraphics[width=0.24\textwidth]{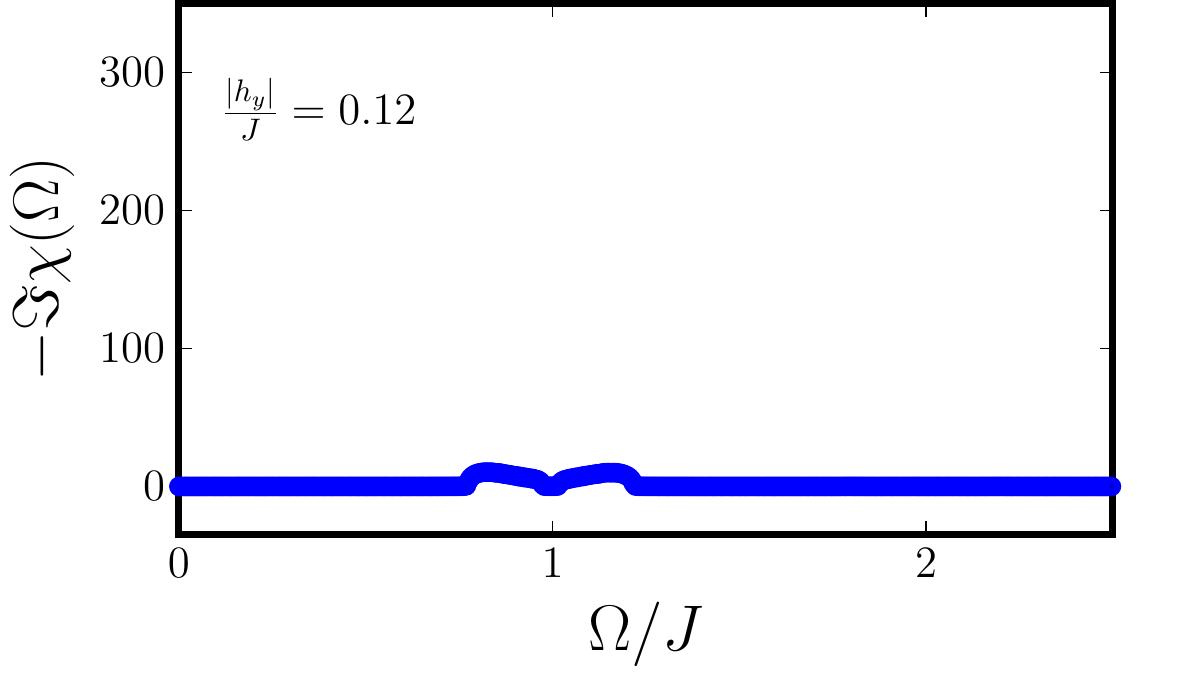}}
\caption{ \label{fig:im_com}
Imaginary part of the spin correlator $\chi (\Omega)$, Eq.~\ref{eq:chi_w}, which is proportional to the second derivative of the charge noise spectrum, Eq.~\eqref{d2sc}, measured across the normal metal film. Parameters used are $\jn/\jp=0.01$, $\dm/\jp=\gm/\jp=0.1$. A phenomenological Lorentzian broadening with width $\delta / J=10^{-3}$ is used to account for scattering and impurity effects. Here, we have set $\hbar =1$, such that both energies and frequencies are measured in units of $J$. The topological quantum paramagnet occurs for $|\hhy|/J<0.1$.}
\end{figure}
%%%%%%%%%%%%%%%%%%%%%%%%%%%%%%%%%

For the spin current noise spectral density, Eq.~\eqref{noise_spectral_density}, we need the correlators on each leg. In short, we need the Fourier transform of the following correlator at the end site of the ladder
\begin{equation}
\label{eq:corr_t}
\sum_{m=1,2} \langle \hS^{x}_{m1} (t) \hS^{x}_{m1} (0) + \hS^{z}_{m1} (t) \hS^{z}_{m1} (0) \rangle \ ,
\end{equation}
where $1$ denotes the left end site. 
In terms of the triplon operators, the required correlators take the form
\begin{align}
\label{eq:sx_sx}
&\sum_{m=1,2} \hS^{x}_{m} (t) \hS^{x}_{m} (0)  \nonumber \\
&= \frac{1}{2} \left[ - \txd (t) \txd (0) + \txd (t) \tx (0) 
+ \tx (t) \txd (0) - \tx (t) \tx (0) \right]  \nonumber \\
&+ \frac{1}{2} \bigg[ - \tyd (t) \tz (t) \tyd (0) \tz (0)  + \tyd (t) \tz (t) \tzd (0) \ty (0)  \nonumber \\
&+ \tzd (t) \ty (t) \tyd (0) \tz (0) - \tzd (t) \ty (t) \tzd (0) \ty (0) \bigg] \,,  
\end{align}
and
\begin{align}
%%%%%%%
\label{eq:sz_sz}
&\sum_{m=1,2} \hS^{z}_{m} (t) \hS^{z}_{m} (0)  \nonumber \\
&= \frac{1}{2} \left[ - \tzd (t) \tzd (0) + \tzd (t) \tz (0) 
+ \tz (t) \tzd (0) - \tz (t) \tz (0) \right]  \nonumber \\
&+ \frac{1}{2} \bigg[ - \tyd (t) \tx (t) \tyd (0) \tx (0)  + \tyd (t) \tx (t) \txd (0) \ty (0)  \nonumber \\
&+ \txd (t) \ty (t) \tyd (0) \tx (0) - \txd (t) \ty (t) \txd (0) \ty (0) \bigg] \,.
\end{align}
where, for brevity, we have suppressed the site index $i=1$ of the spin and triplon operators. The first line in Eqs.~\eqref{eq:sx_sx} and \eqref{eq:sz_sz} contribute to the single-particle response, while the second and third lines give the two-particle response. Hence, the second derivative of the charge noise spectrum, Eq.~\eqref{d2sc}, is proportional to
\begin{equation}
\label{eq:chi_w} 
\chi (\Omega) \equiv\sum_{m=1,2} \left[ \chi^{xx}_{m1,m1} (\Omega) + \chi^{zz}_{m1,m1} (\Omega) \right] \, ,
\end{equation}
which is obtained from the Fourier transforms of Eqs.~\eqref{eq:sx_sx} and~\ref{eq:sz_sz}, see SM for details~\cite{supp_info}.

\begin{figure}
\centering
\subfloat{\includegraphics[width=0.24\textwidth]{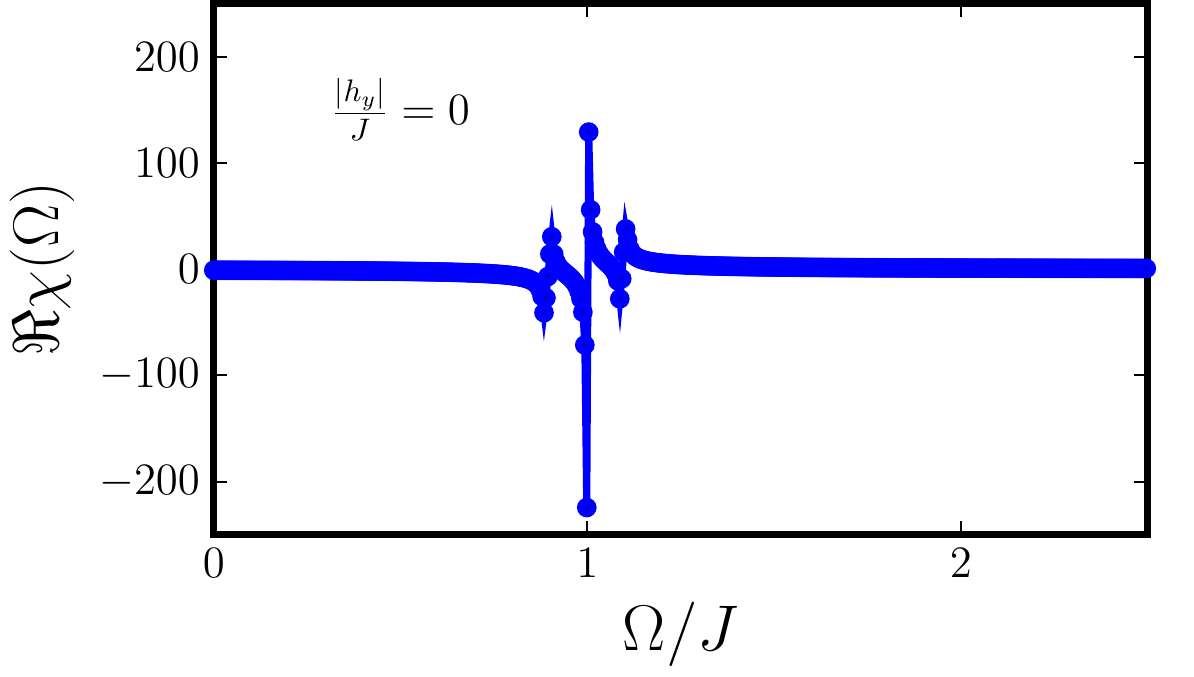}} ~~
\subfloat{\includegraphics[width=0.24\textwidth]{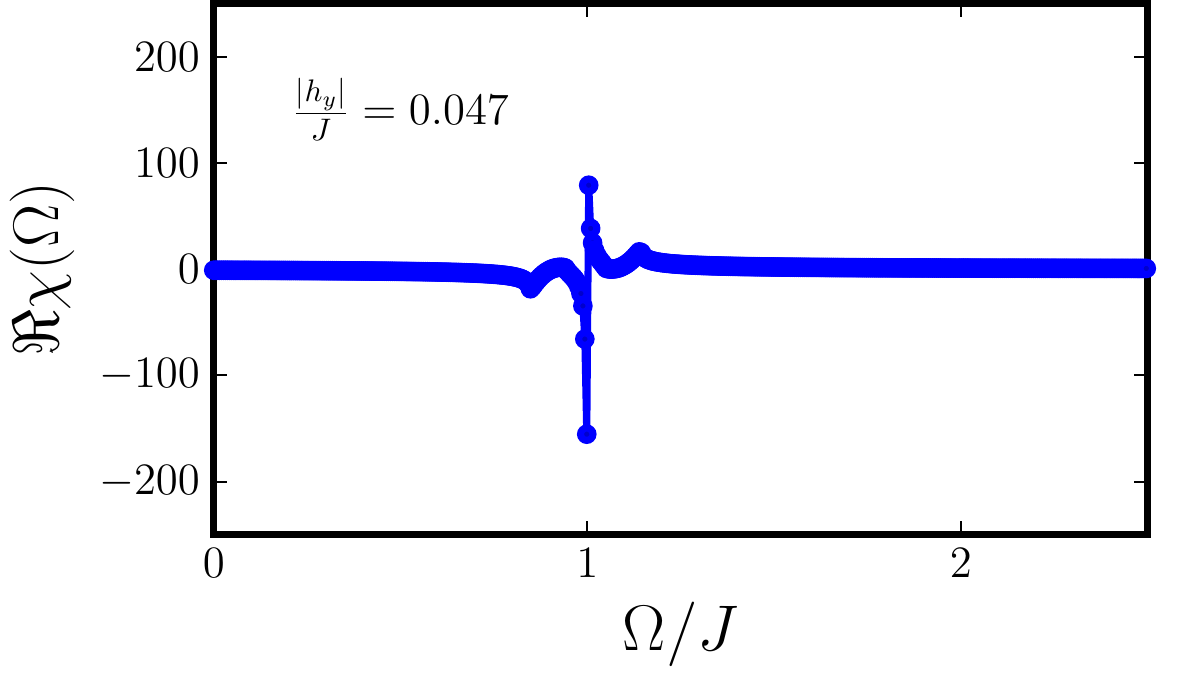}} \\
\subfloat{\includegraphics[width=0.24\textwidth]{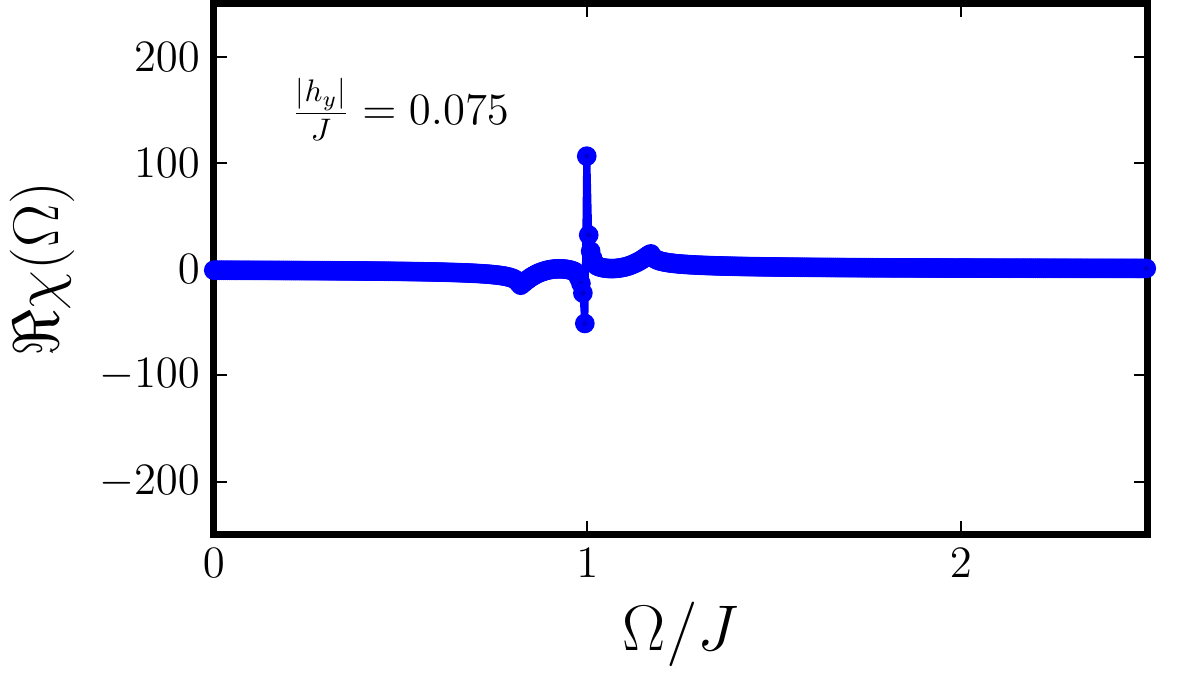}} ~~
\subfloat{\includegraphics[width=0.24\textwidth]{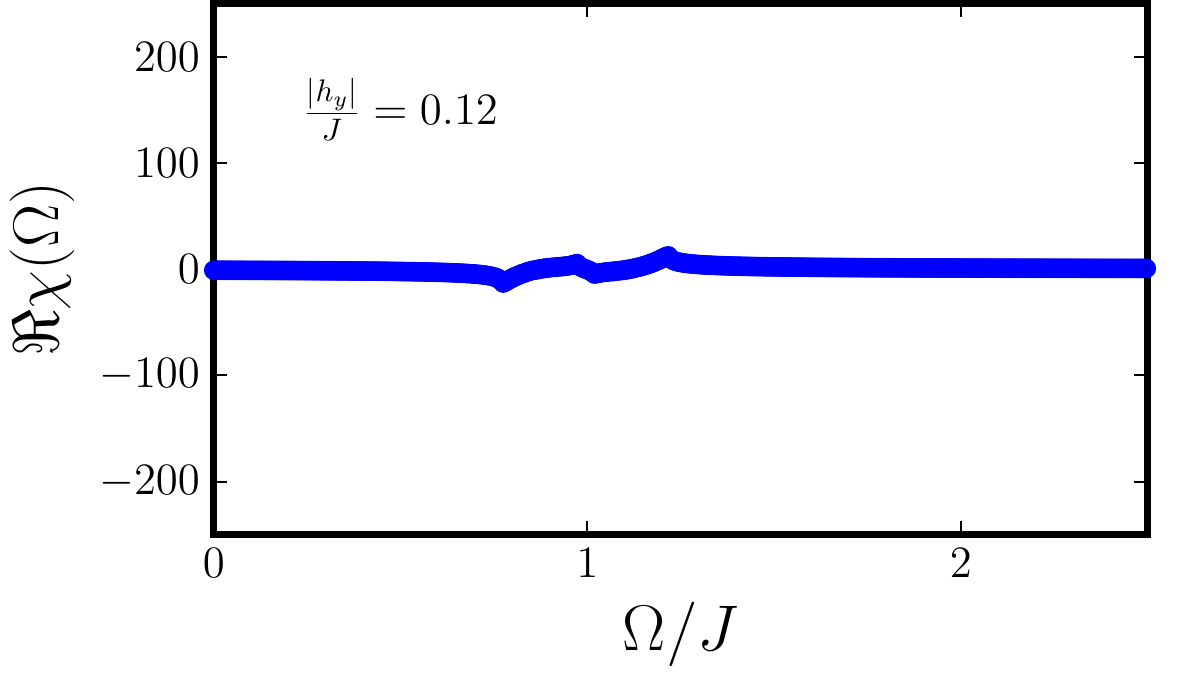}}
\caption{\label{fig:re_com} Real part of the spin correlator $\chi (\Omega)$,  Eq.~\eqref{eq:chi_w}, for the same parameters as in Fig.~\ref{fig:im_com}.
The topological quantum paramagnet occurs for $|\hhy|/J<0.1$.}
\end{figure}

As is evident, $\x(\W)$ has both real and imaginary parts. Due to the factor of $\ii$, the real part of the SHNS observable, Eq.~\eqref{d2sc}, is proportional to  $-\Im\{\chi\}$, while its imaginary part is proportional to $\Re\{\chi\}$. These are plotted in Figs.~\ref{fig:im_com} and~\ref{fig:re_com}, respectively. Within the topological phase, i.e., the topological quantum paramagnet, which occurs for $| h_y | < 0.1$, there are {\em in-gap} localized end states around $\Omega / J =1 $. (Here we set $\hbar = 1$, such that both energies and frequencies are measured in units of $J$.) Hence, we expect sharp peaks in the imaginary part $-\Im\{\chi\}$, as is seen in Fig.~\ref{fig:im_com}. Correspondingly, the real part $\Re\{\chi\}$ exhibits $1/\Omega$ singularities at $\Omega / J =1 $. Such peaks are absent in the topologically trivial paramagnetic phase, due to the absence of the localized end states. The appearance of these peaks provide a clear distinguishing feature to identify the topological quantum paramagnetic phase using SHNS. We note that the dominant features in $\chi ( \Omega)$ at $\Omega / J =1$ originate from the single-particle response, while the contributions from the two-particle response are negligibly small.

% there are also contributions
%from the two-particle response, which are visible near  $\Omega / J =2$ (insets of Figs.~\ref{fig:im_com} and~\ref{fig:re_com}).
%These two-particle contributions are combinations of the $y$-mode with either the $x$- or the $z$-mode [see Eqs.~\eqref{eq:sx_sx} and~\eqref{eq:sz_sz}].
%Since the $y$-mode has energies close to $\Omega / J=1$,} it can combine with the end-state 
%energy in the topological phase to give peaks around $\Omega / J=2$. Other peaks in the imaginary part are coming from the {\em bulk} triplon bands. 

%%%%%%%%%%%%%%%%%%%%%%%%%%%%%%%%%%%%%% Real-part plots (combo)
%%%%%%%%%%%%%%%%%%%%%%%%%%%%%%%%%%%%%

We now estimate the magnitude of the predicted charge noise effect for Pt thin films deposited on top of a spin-ladder material. The charge current noise can be measured as voltage noise if the normal metal is electrically open. Here, we consider a Pt film of thickness $d=7$~nm attached laterally to the ends of a stack of the quantum spin ladder material BiCu$_2$PO$_6$~\cite{tsirlinPRB10,wangJCG10,kohamaPRL12} with the coupling interfacial area of 5~mm~$\times$~1~mm. BiCu$_2$PO$_6$ has recently been identified as a promising candidate for realizing the topological triplon phase~\cite{joshiPRB17}. Converting Eq.~\eqref{d2sc} to voltage fluctuations, 
\beq
\frac{d^2S^z_V(\W)}{d\W^2}=2 \ii\square{\Theta\frac{2e}{\hbar}\sN\rho\ell\frac{\la}{d}\tanh\round{\frac{d}{2\la}}\round{\frac{\eta a^3mk_F}{2\p^2\hbar}}}^2\x(\W)\ ,
\eeq
where $\rho$ is the resistivity of Pt and $\ell$ is the length over which the voltage drop is measured (i.e., $\ell=5$~mm here). To obtain the derivatives of the spectrum, one needs to measure the noise over a frequency window of width, e.g., $0.1J/\hbar$, straddling $\W\sim J/\hbar$. Using effective Pt electron mass of $m\approx13m_e$~\cite{kasapBOOK06}, $\rho\approx10^{-8}~\W$m, spin Hall angle of $\Theta\approx0.1$, spin diffusion length of $\la\approx3$~nm~\cite{liuPRL11}, lattice constant for Pt of $a\approx4\AA\sim k_F^{-1}$, $\sN\approx(8\AA)^{-2}$ based on the crystal structure for BiCu$_2$PO$_6$~\cite{tsirlinPRB10}, interfacial exchange constant of $\eta/k_B\approx1$~K, and maximum peak height of $100J/\hbar$ for $\chi(\W)$ (see Figs.~\ref{fig:im_com} and \ref{fig:re_com}), the voltage noise amplitude becomes of order $S_V\sim10^{-21}$~V$^2$/Hz. Voltage fluctuations of this magnitude have been detected in~\cite{kamraPRB14}. Finally, we note from Figs.~\ref{fig:im_com} and \ref{fig:re_com} that signatures of the topological end-states appear at a frequency scale of the order of $J$, which in the spin ladder materials is of the order of   $J\sim 1$--$10$~meV $\sim100$--$1000$~GHz~\cite{wangJCG10}. These frequency scales are accessible given the availability of high-frequency noise spectral analyzers with ranges up to a few hundred GHz~\cite{kochPRB82,*onacPRL06}.

Electric noise in the metal may have other contributions, such as Johnson-Nyquist and $1/f$ noises, that may mask the noise generated by the quantum paramagnet. We do not expect $1/f$ noise to be problematic in the high frequency range where the topological signatures are predicted to arise. Furthermore, since any sources of noise that are independent of frequency (e.g., thermal white noise) or vary linearly over the relevant frequency range are eliminated once the second derivative of the noise spectrum is taken, SHNS should be capable of exposing the spin noise computed in this work.
 
In conclusion, we studied the equilibrium charge noise in a metal in contact with a quantum paramagnet.  We showed that the interfacial spin fluctuations of the paramagnet induce spin current fluctuations in the metal, which via the inverse spin Hall effect (ISHE) lead to charge fluctuations. More precisely, the second derivative of the charge noise spectrum in the metal is directly proportional to the dynamical spin correlation function of the quantum paramagnet. Hence, by measuring the ISHE-induced charge fluctuations it is possible to probe the dynamical spin correlation of any quantum paramagnet in a table-top experiment. This is particularly useful to detect edge states of topological quantum paramagnets, as these are quite difficult to observe with other probes. We have demonstrated this for the case of a topologically nontrivial quantum spin ladder, whose topological edge states lead to distinct features in the charge noise at a frequency of the order of $J$. We expect that the discussed detection technique can be   applied to probe the topological edge states of a wider class of quantum magnets, such as, the spin-1/2 edge states of  the AKLT  chain~\cite{affleckPRL87}, or the chiral magnon edge states of compounds described by the Kitaev-Heisenberg model~\cite{mcclartyCM18,joshiCM18}. We hope that our findings will stimulate experimental investigations along these lines.

%%%%%%%%%%%%%%%%%%%%%%%%%%%%%%%%%%%%%%%%%%%%%%%%%%%%%%%%%%%%%%%%%
\newpage
\begin{appendices}
\section{Supplemental Material: \\ I.~Derivation of charge current spectral noise}

In this section we derive Eqs.~(5) and (7) from the main text. The normal metal is modeled as a free electron gas with the usual quadratic dispersion $\ve_\bk=\hbar^2k^2/2m$, temperature $T$ and chemical potential $\mu$. Then the spectrum of spin current fluctuations polarized along the $y$ axis [denoted $S^y_s(\W)$ in the main text] can be computed perturbatively to lowest non-trivial order in $\eta$ as
%\beq
%\hH_e=\sum_\s\int_{y<0} d^3\bx\ \psi^\dag_\s(\bx)\round{-\frac{\hbar^2\nabla^2}{2m}}\psi_\s(\bx)\ .
%\eeq
%obeying the Fermi-Dirac distribution function $\langle\psi^\dag_{\bk\s}\psi_{\bk\s}\rangle=(2\p)^3\de_{\s\s'}\de(\bk-\bk')f_{\bk}$ with
%\beq
%f_{\bk}=\frac{1}{e^{\be\ve_{\bk}}+1}\ ,
%\eeq
%We now calculate the nonequilibrium expectation value of the noise spectral density $S_s(\W)$ using the Keldysh formalism. To lowest non-trivial order in $\eta$, we obtain
\begin{multline}
S^y_s(\W)=\round{\frac{\eta a^3}{2}}^2\int dte^{-i\W t}\\
\times\square{\ang{\hT_y(0)\hT_y^\dag(t)}_0+\ang{\hT_y^\dag(0)\hT_y(t)}_0}\ ,
\end{multline}
where $\hT_y$ was given in the main text and subscript 0 denotes equilibrium thermal averages. We introduce a Fourier transform for the electron field that respects the finite boundaries in the $x$ direction, i.e.,
\beq
\hat\psi_\s(\bx)=\frac{1}{\sqrt{\sA_i}}\sqrt{\frac{2}{d}}\sum_\bk e^{ik_yy+ik_zz}\cos(k_xx)\hc_{\bk\s}\ ,
\eeq
where $(k_y,k_z)=(2\p/L)(n_y,n_z)$ with $n_y,n_z\in\mathbb{Z}$ and $k_x=n_x\p/L$ with $n_x=1,2,3,\dots$. Then upon insertion of $\hT_y$ into the expression, we obtain
\begin{multline}
S^y_s(\W)=2\ii(\eta a^3)^2\int\frac{d\nu}{2\p}\int\frac{d\w}{2\p}\int\frac{d^3\bk}{(2\p)^3}\int\frac{d^3\bk'}{(2\p)^3}\sum_{\bi,\bj}\\
\de_{i_x,1}\de_{j_x,1}A_\bk(\w+\nu-\W)A_{\bk'}(\w)f(\w+\nu-\W)[1-f(\w)]\\
\times e^{-i(\bk_\perp-\bk'_\perp)\cdot(\bR_\bi-\bR_\bj)}\square{\x^{xx}_{\bi\bj}(\nu)+\x^{zz}_{\bi\bj}(\nu)}\ ,
\end{multline}
where $A_\bk(\w)=2\p\de(\w-\ve_\bk/\hbar)$ is the electron spectral function, $f(\w)=[e^{\be(\hbar\w-\mu)}+1]^{-1}$ is the Fermi-Dirac distribution function, $\bk_\perp=(k_y,k_z)$ collects the transverse wavevectors and the correlators $\chi^{\al\al}_{\bi\bj}(\nu)$ have been defined in the main text. Assuming that the electronic density of states do not vary appreciably for energies near the chemical potential, we may approximate the above integrals as
\begin{multline}
S^y_s(\W)\approx2\ii(\eta a^3)^2\int\frac{d\nu}{(2\p)^2}\int\frac{d^3\bk}{(2\p)^3}\int\frac{d^3\bk'}{(2\p)^3}\sum_{\bi,\bj}\\
\de_{i_x,1}\de_{j_x,1}A_\bk(\mu)A_{\bk'}(\mu)\frac{\W-\nu}{1-e^{-\be\hbar(\W-\nu)}}\\
\times e^{-i(\bk_\perp-\bk'_\perp)\cdot(\bR_\bi-\bR_\bj)}\square{\x^{xx}_{\bi\bj}(\nu)+\x^{zz}_{\bi\bj}(\nu)}\ .
\end{multline}
Performing the integrals over $\bk$ and $\bk'$ immediately gives Eq.~(5). 

In the limit of large Fermi wavevector $k_F$, the square of the sinc-function becomes essentially nonzero only for $\bi=\bj$. So Eq.~(5) can be further approximated as
\begin{multline}
S^y_s(\W)\approx2\ii\round{\frac{\eta a^3mk_F}{2\p^2\hbar}}^2\int d\nu\sum_{\bi}\\
\de_{i_x,1}\square{\x^{xx}_{\bi\bi}(\nu)+\x^{zz}_{\bi\bi}(\nu)}\frac{\W-\nu}{1-e^{-\be\hbar(\W-\nu)}}\ .
\end{multline}
Taking the second derivative of the noise with respect to the spectral frequency $\W$, we obtain
\begin{multline}
S''_s(\W)=2\ii\round{\frac{\eta a_e^3mk_F}{2\p^2\hbar}}^2\int d\nu\sum_{\bi}\de_{i_x,1}\\
\times\square{\x^{xx}_{\bi\bi}(\nu)+\x^{zz}_{\bi\bi}(\nu)}\frac{d^2}{d\W^2}\round{\frac{\W-\nu}{1-e^{-\be\hbar(\W-\nu)}}}\ .
\end{multline}
At low temperatures, the last second derivative factor is strongly peaked only at $\W=\nu$, so we may approximate the integral as
\[
S''_s(\W)=2\ii\round{\frac{\eta a_e^3mk_F}{2\p^2\hbar}}^2\sum_{\bi}\de_{i_x,1}\square{\x^{xx}_{\bi\bi}(\nu)+\x^{zz}_{\bi\bi}(\nu)}\ ,
\]
which, together with Eq.~(4) in the main text, gives Eq.~(7).

\section{II.~Spin correlation functions}

As mentioned in the main text, in the quantum paramagnetic phase we represent spins via bosonic quasiparticles, {\em triplons}, described within the bond-operator theory \cite{sachdevPRB90,romhanyiPRB11}.
Within the harmonic approximation, i.e. neglecting triplon interaction terms, the Hamiltonian of the quantum spin ladder (Eq. (9) in the main text) takes the following form~\cite{joshiPRB17}:
\begin{align}
\label{eq:ham_har}
&\mathcal{H}_{2} = \jp \sum_{i \alpha} \hat{t}^{\dagger}_{i\alpha} \hat{t}_{i\alpha}  %\nonumber \\
+ \frac{\jn}{2} \sum_{i\alpha} \big[ \hat{t}^{\dagger}_{i\alpha} \hat{t}_{i+1 \alpha}
- \hat{t}^{\dagger}_{i\alpha} \hat{t}^{\dagger}_{i+1\alpha}  + H.c. \big] \nonumber \\
&+\frac{\dm}{2} \sum_{i} \big[ \hat{t}^{\dagger}_{iz} \hat{t}_{i+1 x} - \hat{t}^{\dagger}_{iz} \hat{t}^{\dagger}_{i+1x}  
- \hat{t}^{\dagger}_{ix} \hat{t}_{i+1 z} + \hat{t}^{\dagger}_{ix} \hat{t}^{\dagger}_{i+1z}  + H.c. \big] \nonumber \\
&+\frac{\gm}{2} \sum_{i} \big[ \hat{t}^{\dagger}_{iz} \hat{t}_{i+1 x} - \hat{t}^{\dagger}_{iz} \hat{t}^{\dagger}_{i+1x}  
+ \hat{t}^{\dagger}_{ix} \hat{t}_{i+1 z} - \hat{t}^{\dagger}_{ix} \hat{t}^{\dagger}_{i+1z} + H.c. \big] \nonumber \\
&+\ii \hy \sum_{i} \big[ \hat{t}^{\dagger}_{ix} \hat{t}_{iz} - \hat{t}^{\dagger}_{iz} \hat{t}_{ix} \big] \,, \\
%%%%
\label{eq:m_def}
&= \Psi^{\dagger} \mathcal{M}_{2} \Psi \,,
\end{align}
where $\Psi=\left( \hat{t}_{1x} \ldots \hat{t}_{Nx}, \hat{t}_{1z} \ldots \hat{t}_{Nz}, 
\hat{t}^{\dagger}_{1x} \ldots \hat{t}^{\dagger}_{Nx}, \hat{t}^{\dagger}_{1z} \ldots \hat{t}^{\dagger}_{Nz} \right)^{T}$.

The eigenmodes of the above triplon Hamiltonian are 
obtained by diagonalizing the non-Hermitian matrix $\Sigma \mathcal{M}_{2}$, where $\Sigma=$diag($1,1,-1,-1$). This is done in the following way
\cite{blaizotBOOK85,muccioloPRB04,wesselPRB05}:
\begin{equation}
\label{eq:diag_ham}
\Omega=T^{\dagger} \mathcal{M}_{2} T ,
\end{equation}
where $\Omega$ is a diagonal matrix, which contains eigenmodes of the Hamiltonian, and 
\begin{equation}
\label{eq:t_def}
T=
\begin{bmatrix}
U & V \\
V^{*} & U^{*} 
\end{bmatrix}
\end{equation}
is the $4N \times 4N$ transformation matrix. It satisfies the condition
\begin{equation}
\label{eq:t_rule}
T^{\dagger} \Sigma T = T \Sigma T^{\dagger} = \Sigma \,,
\end{equation}
which ensures that the bosonic commutation relations are satisfied. 
The $2N \times 2N$ matrices $U$ and $V$ contain the Bogoliubov coefficients, which are related to the right eigenvector of $\Sigma \mathcal{M}_{2}$, such that for an eigenfrequency $\omega_n$,
\begin{equation}
\label{eq:uv}
\Sigma \mathcal{M}_{2} |\phi_n \rangle = \omega_n |\phi_n \rangle \,; ~~~~
|\phi_n \rangle = 
\begin{bmatrix}
u_n \\
v^{*}_{n}
\end{bmatrix} \,.
\end{equation} 
Since we are interested in the edge phenomena, the eigenmodes are to be calculated for a ladder with open boundary condition. 

In order to calculate the correlator $\chi(\Omega)$ (Eq. (14) in the main text), we therefore represent the triplon operators ($\hat{t}$) in terms of the Bogoliubov quasiparticles ($\ta$) as follows:
\begin{align}
\label{eq:t1x}
\hat{t}_{1x} &= \sum_{m=1}^{N} \bigg[ u_{1,m} \ta_{mx} + u_{1,m+N} \ta_{mz}  \nonumber \\
&~~~~~~~~+ v_{1,m} \ta^{\dagger}_{mx} + v_{1,m+N} \ta^{\dagger}_{mz} \bigg] \,, \\
%%%%%
\label{eq:t1z}
\hat{t}_{1z} &= \sum_{m=1}^{N} \bigg[ u_{1+N,m} \ta_{mx} + u_{1+N,m+N} \ta_{mz}  \nonumber \\
&~~~~~~~~+ v_{1+N,m} \ta^{\dagger}_{mx} + v_{1+N,m+N} \ta^{\dagger}_{mz} \bigg] \,, \\
%%%%%%
\label{eq:t1y}
\hat{t}_{1y} &= \sum_{m=1}^{N} \bigg[ \uy_{1,m} \ta_{my} + \vy_{1,m} \ta^{\dagger}_{my} \bigg] \,.
\end{align}
Here, $u,v,\uy,\vy$ are Bogoliubov coefficients obeying standard rules in Eq. \ref{eq:t_rule}. Since the $t_y$ mode is decoupled at the harmonic level, in quartic order in $\ta$ only terms of the following form will give non-zero contributions once we substitute Eqs. \ref{eq:t1x} - \ref{eq:t1y} into Eqs. (5) and (6) in the main text:
$\ta_{y} \ta_{z} \ta^{\dagger}_{y} \ta^{\dagger}_{z}$, $\ta_{y} \ta_{x} \ta^{\dagger}_{y} \ta^{\dagger}_{x}$, 
$\ta_{x} \ta_{z} \ta^{\dagger}_{x} \ta^{\dagger}_{z}$, $\ta_{x} \ta_{z} \ta^{\dagger}_{z} \ta^{\dagger}_{x}$, 
$\ta_{z} \ta_{x} \ta^{\dagger}_{x} \ta^{\dagger}_{z}$, $\ta_{z} \ta_{x} \ta^{\dagger}_{z} \ta^{\dagger}_{x}$, 
$\ta_{x} \ta_{x} \ta^{\dagger}_{x} \ta^{\dagger}_{x}$ and $\ta_{z} \ta_{z} \ta^{\dagger}_{z} \ta^{\dagger}_{z}$.

At the harmonic level, $\ta(t) = e^{\ii \mathcal{H}_{2} t} \ta e^{-\ii \mathcal{H}_{2} t}$. It is then straight forward to evaluate the correlators in Eqs. (12) and (13) in the main text and then Fourier transform to obtain the frequency dependence. 
Therefore, for the correlator arising from $\hat{S}^{x}$ we obtain,
%\begin{equation}
%\label{eq:chi_w}
%\chi (\Omega) \equiv \sum_{m=1,2} \left[ \chi^{xx}_{11,m} (\Omega) + \chi^{zz}_{11,m} (\Omega) \right] \,,
%\end{equation}
\begin{align}
\label{eq:chi_xx}
&\sum_{m=1,2} \chi^{xx}_{m1,m1} (\Omega) \nonumber \\
&= \frac{1}{2} \sum^{N}_{m=1} \frac{1}{\Omega - \omega_{mx}}
\big[ - v^{*}_{1,m} u^{*}_{1,m} + v^{*}_{1,m} v_{1,m}  \nonumber \\
&+ u_{1,m} u^{*}_{1,m} - u_{1,m} v_{1,m}  %\nonumber \\
%&~~~~~~~~
- v^{*}_{1+N,m} u^{*}_{1+N,m}  \nonumber \\
&+ v^{*}_{1+N,m} v_{1+N,m}
+ u_{1+N,m} u^{*}_{1+N,m} - u_{1+N,m} v_{1+N,m}
\big] \nonumber \\
&+ \frac{1}{2}  \sum_{m,n=1}^{N} \frac{1}{\Omega - (\omega^{y}_{m} + \omega^{x}_{n})} 
\bigg[ -\uy^{*}_{1,m} \vy^{*}_{1,m} u_{1+N,n} v_{1+N,n} \nonumber \\
&+ |\vy_{1,m}|^{2} |u_{1+N,n}|^{2} %\nonumber \\
%&~~~~~~~~
+ |\uy_{1,m}|^{2} |v_{1+N,n}|^{2} \nonumber \\
&-\uy_{1,m} \vy_{1,m} u^{*}_{1+N,n} v^{*}_{1+N,n}  %\nonumber \\
%&~~~~~~~~~~~~~~~~~~~~~~
-\uy^{*}_{1,m} \vy^{*}_{1,m} u_{1,n} v_{1,n} \nonumber \\
&+ |\vy_{1,m}|^{2} |u_{1,n}|^{2} %\nonumber \\
%&~~~~~~~~
+ |\uy_{1,m}|^{2} |v_{1,n}|^{2}  \nonumber \\
&-\uy_{1,m} \vy_{1,m} u^{*}_{1,n} v^{*}_{1,n}
\bigg] \,.
\end{align}
Similarly, the correlator arising from $\hat{S}^{z}$ is given by
\begin{align}
\label{eq:chi_zz}
&\sum_{m=1,2} \chi^{zz}_{m1,m1} (\Omega)  \nonumber \\
&=\frac{1}{2} \sum^{N}_{m=1} \frac{1}{\Omega - \omega_{mz}} 
\big[ - v^{*}_{1,m+N} u^{*}_{1,m+N} + v^{*}_{1,m+N} v_{1,m+N}  \nonumber \\
&+ u_{1,m+N} u^{*}_{1,m+N} - u_{1,m+N} v_{1,m+N}  \nonumber \\
%&~~~~~~~~
&- v^{*}_{1+N,m+N} u^{*}_{1+N,m+N}  %\nonumber \\
+ v^{*}_{1+N,m+N} v_{1+N,m+N}  \nonumber \\
&+ u_{1+N,m+N} u^{*}_{1+N,m+N} - u_{1+N,m+N} v_{1+N,m+N}
\big]   \nonumber \\
%%%%%
&+ \frac{1}{2}  \sum_{m,n=1}^{N} 
\frac{1}{\Omega - (\omega^{y}_{m} + \omega^{z}_{n})} 
\bigg[ -\uy^{*}_{1,m} \vy^{*}_{1,m} u_{1+N,n+N} v_{1+N,n+N}  \nonumber \\
&+ |\vy_{1,m}|^{2} |u_{1+N,n+N}|^{2}  %\nonumber \\
%&~~~~~~~~
+ |\uy_{1,m}|^{2} |v_{1+N,n+N}|^{2}  \nonumber \\
&-\uy_{1,m} \vy_{1,m} u^{*}_{1+N,n+N} v^{*}_{1+N,n+N}  %\nonumber \\
%&~~~~~~~~~~~~~~~~~~~~~~
-\uy^{*}_{1,m} \vy^{*}_{1,m} u_{1,n+N} v_{1,n+N} \nonumber \\
&+ |\vy_{1,m}|^{2} |u_{1,n+N}|^{2}  %\nonumber \\
%&~~~~~~~~
+ |\uy_{1,m}|^{2} |v_{1,n+N}|^{2}  \nonumber \\
&-\uy_{1,m} \vy_{1,m} u^{*}_{1,n+N} v^{*}_{1,n+N}
\bigg]  \,.
\end{align}
In both the above equations, the first expression on the RHS gives the single-particle contribution, while the second term gives the 
two-particle contribution. The required correlator $\chi (\Omega)$ [Eq. (14)] is then obtained as a sum of Eq. \ref{eq:chi_xx} and \ref{eq:chi_zz}.

\end{appendices}


%\bibliography{/Users/so/Dropbox/refs/ref.bib}
%\bibliography{ref.bib}
\end{document}